\documentclass{Rinton-P9x6}

\newdimen\snglfig
\newdimen\dblfig
\dblfig=\textwidth \advance\dblfig by-5pt \divide\dblfig by 2\relax

\def\etmiss{\slashchar{E}_T}
\def\tg{{\tilde g}}
\def\tq{{\tilde q}}
\def\tchi{{\tilde\chi}}

\def\lsp{{\tilde\chi_1^0}}
\def\tG{{\tilde G}}
\def\ttau{{\tilde\tau}}
\def\tell{{\tilde\ell}}
\def\tn{{\tilde\nu}}

\def\GeV{{\rm GeV}}
\def\TeV{{\rm TeV}}
\def\Meff{M_{\rm eff}}

\def\sgn{\mathop{\rm sgn}}

\def\fb{{\rm fb}}
\def\fbi{{\rm fb^{-1}}}

\def\mhalf{{m_{1/2}}}

\def\bentarrow{\:\raisebox{1.1ex}{\rlap{$\vert$}}\!\kern-.3pt\rightarrow}

\def\dofig#1#2{\epsfxsize=#1\centerline{\epsfbox{#2}}}
\def\dofigs#1#2#3{\centerline{\epsfxsize=#1\epsfbox{#2}%
   \hfil\epsfxsize=#1\epsfbox{#3}}}

\def\simge{
    \mathrel{\rlap{\raise 0.511ex
        \hbox{$>$}}{\lower 0.511ex \hbox{$\sim$}}}}
\def\simle{
    \mathrel{\rlap{\raise 0.511ex 
        \hbox{$<$}}{\lower 0.511ex \hbox{$\sim$}}}}

\def\slashchar#1{\setbox0=\hbox{$#1$}           
   \dimen0=\wd0                                 
   \setbox1=\hbox{/} \dimen1=\wd1               
   \ifdim\dimen0>\dimen1                        
      \rlap{\hbox to \dimen0{\hfil/\hfil}}      
      #1                                        
   \else                                        
      \rlap{\hbox to \dimen1{\hfil$#1$\hfil}}   
      /                                         
   \fi}                                         %

\catcode`@=11
\def\citenum#1{\csname b@#1\endcsname}
\catcode`@=12

\begin{document}

\font\twelvess=cmss10 scaled \magstep1

\begingroup
\parindent=20pt
\thispagestyle{empty}
\vbox to 0pt{
\dimen9=\textheight 
\advance\dimen9 by -8.9in 
\divide\dimen9 by 2 
\vskip\dimen9
\dimen8=\textwidth
\advance\dimen8 by -6.5in
\divide\dimen8 by 2

\moveleft-\dimen8\vbox to 8.9in{\hsize=6.5in

\centerline{\twelvess BROOKHAVEN NATIONAL LABORATORY}
\vskip6pt
\hrule
\vskip1pt
\hrule
\vskip4pt
\hbox to \hsize{July, 2003 \hfil BNL-HET-03/16}
\vskip3pt
\hrule
\vskip1pt
\hrule
\vskip3pt

\vskip1.5in
\centerline{\LARGE\bf SUSY Signatures in ATLAS at LHC}
\vskip.5in
\centerline{\bf Frank E. Paige}
\vskip4pt
\centerline{Physics Department}
\centerline{Brookhaven National Laboratory}
\centerline{Upton, NY 11973 USA}

\vskip.75in

\centerline{ABSTRACT}

\vskip8pt
\narrower\narrower
This talk summarizes work by the ATLAS Collaboration at the
CERN Large Hadron Collider on the search SUSY particles and Higgs bosons
and on possible measurements of their properties.

\vskip1in

Invited talk at {\sl SUGRA 20: 20 Years of SUGRA and the Search for SUSY
and Unification} (Northeastern University, Boston, 17--21~March, 2003.)

\vskip0pt
\vfil\footnotesize
        This manuscript has been authored under contract number
DE-AC02-98CH10886 with the U.S. Department of Energy.  Accordingly,
the U.S.  Government retains a non-exclusive, royalty-free license to
publish or reproduce the published form of this contribution, or allow
others to do so, for U.S. Government purposes.

\vskip0pt} 
\vss} 

\newpage
\thispagestyle{empty}
\hbox{\ }
\newpage
\setcounter{page}{1}


\title{SUSY Signatures in ATLAS at LHC}

\author{Frank E. Paige}

\address{Brookhaven National Laboratory, Upton, NY 11973
}

\maketitle

\abstracts{This talk summarizes work by the ATLAS Collaboration at the
CERN Large Hadron Collider on the search SUSY particles and Higgs bosons
and on possible measurements of their properties.}

\section{Introduction}

It has been twenty years since Richard Arnowitt, Ali Chamseddine, and
Pran Nath introduced minimal supergravity (mSUGRA) as a
phenomenologically viable model of SUSY breaking\cite{mSUGRA}. This talk
summarizes results from the {\sl ATLAS Detector and Physics Performance
TDR}\cite{TDR} and more recent work by the ATLAS Collaboration on the
search for and possible measurements of SUSY particles at the LHC. It
also discusses measurements of Higgs bosons, which are a necessary part
of SUSY. Much of the ATLAS work continues to be based on the mSUGRA
model commemorated at this meeting.

If SUSY exists at the TeV scale, then gluinos and squarks will be
copiously produced at the LHC. Their production cross sections are
comparable to the jet cross section as the same $Q^2$; if $R$ parity is
conserved, they have distinctive decays into jets, leptons, and the
invisible lightest SUSY particle (LSP) $\lsp$, which gives $\etmiss$.
Since ATLAS (and CMS) are designed to detect all of these, simple cuts
can separate SUSY events from the Standard Model (SM) background. The
main problem at the LHC is not to discover SUSY but to make precise
measurements to determine the masses and other properties of SUSY
particles. This will help to understand how SUSY is broken. SUSY models
in which $R$ parity is violated have also been studied,\cite{TDR} but
they will not be discussed here.

Since the main background for SUSY is SUSY, ATLAS has emphasized studies
of specific SUSY model points. Most of these studies start by generating
the signal and the potential SM backgrounds using a parton shower Monte
Carlo (Herwig\cite{herwig}, Isajet\cite{isajet}, or
Pythia\cite{pythia}). The detector response is simulated using
parameterized resolutions and acceptances derived from
GEANT\cite{geant}, and an analysis is developed to isolate specific SUSY
channels. Recently some work has been done using full GEANT simulation
and reconstruction directly.

\begin{figure}[t]
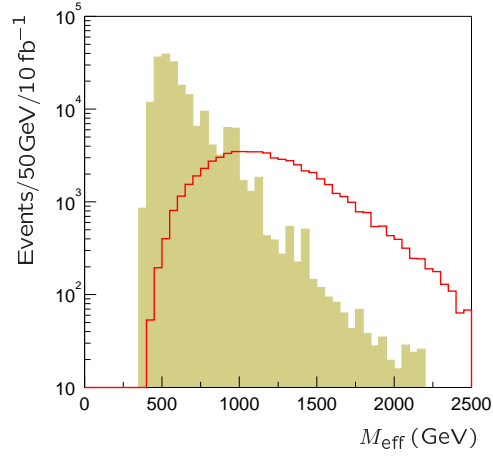

\dofig{2.5in}{c10_meffnew.epsi}
\caption{$\Meff$ distribution for a typical mSUGRA point and SM
backgrounds after cuts. \label{c10_meffnew}}
\end{figure}

\begin{figure}[t]
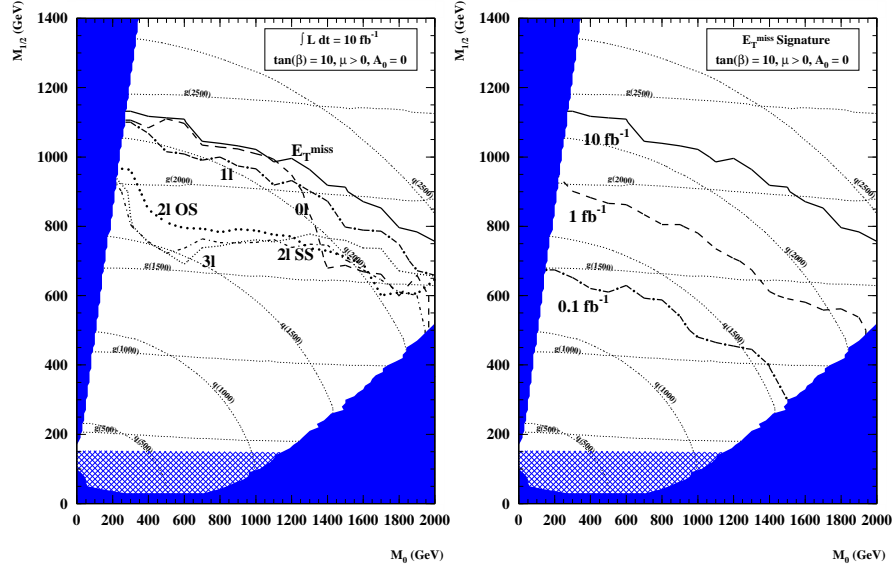

\dofigs{\dblfig}{10gt0p.epsi}{10gt0fb.epsi}
\caption{Search limits for the mSUGRA model in various channels for
$10\,\fbi$ (left) and overall for various luminosities
(right).\protect\cite{tovey} \label{10gt0p}}
\end{figure}

\section{Search for SUSY Particles at LHC}

For masses in the TeV range SUSY production at the LHC is dominated by
$\tg$ and $\tq$. Leptonic decays may or may not be large, but jets and
$\etmiss$ are always produced, and these generally give the best reach.
Consider an mSUGRA with $m_0=100\,\GeV$, $\mhalf=300\,\GeV$, $A_0=0$,
$\tan\beta=10$, $\sgn\mu=+$. Require $\etmiss>100\,\GeV$, at least four
jets with $E_T>100, 50, 50, 50\,\GeV$, and plot as a measure of the
hardness of the collision $\Meff = \etmiss + \sum_j E_{T,j}$.  Then as
Figure~\ref{c10_meffnew} shows the SUSY signal dominates for large
$\Meff$. The search limits from this sort of analysis for mSUGRA
requiring $S>10$ and $S/\sqrt{B}>5$ reach more than $1\,\TeV$ for only
$0.1\,\fbi$ and $2\,\TeV$ for $10\,\fbi$; see Figure~\ref{10gt0p}.

While the AMSB model\cite{amsb} is quite different, the reach in
$M_\tg,M_\tq$ is similar: above $2\,\TeV$ for $100\,\fbi$.  Overall
reach depends mainly on $\sigma(M_\tg,M_\tq)$ provided that $M_\lsp \ll
M_\tg,M_\tq$, so one expects similar reach in most $R$-conserving
models. This should be sufficient if SUSY is related to the naturalness
of the electroweak scale.

\begin{figure}[t]
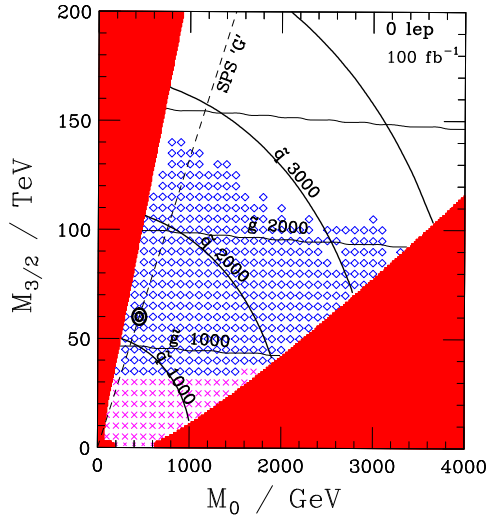

\dofig{\snglfig}{amsbnolep.epsi}
\caption{Search limits for the AMSB model.\protect\cite{barr} 
\label{amsbnolep}}
\end{figure}

\section{SUSY Particle Measurements} 

If $R$ parity is conserved, all SUSY particles decay to an invisible LSP
$\lsp$, so there are no mass peaks. But it is possible to identify
particular decays and to measure their kinematic endpoints, determining
combinations of masses.\cite{hpssy,TDR} The three-body decay $\tchi_2^0
\to \lsp \ell^+\ell^-$ gives a dilepton endpoint at $M_{\ell\ell} =
M_{\tchi_2^0} - M_{\lsp}$, while $\tchi_2^0 \to \tell^\pm \ell^\mp \to
\lsp\ell^+\ell^-$ gives a triangular distribution with an endpoint at
$$
M_{\ell\ell} =
\sqrt{(M_{\tchi_2^0}^2-M_{\tell}^2)(M_{\tell}^2-M_\lsp^2)}/M_\tell\,.
$$
These endpoints can be measured by requiring two isolated leptons in
addition to multijet and $\etmiss$ cuts like those described above. If
lepton flavors are separately conserved, then contributions from two
independent decays cancel in the combination $e^+e^- + \mu^+\mu^- -
e^\pm\mu^\mp$ after acceptance corrections. The resulting distributions
after cuts, Figure~\ref{p4_mll}, are very clean and allow a precise
measurement of the endpoint. The shape allows one to distinguish
two-body and three-body decays.

\begin{figure}[t]
\dofigs{\dblfig}{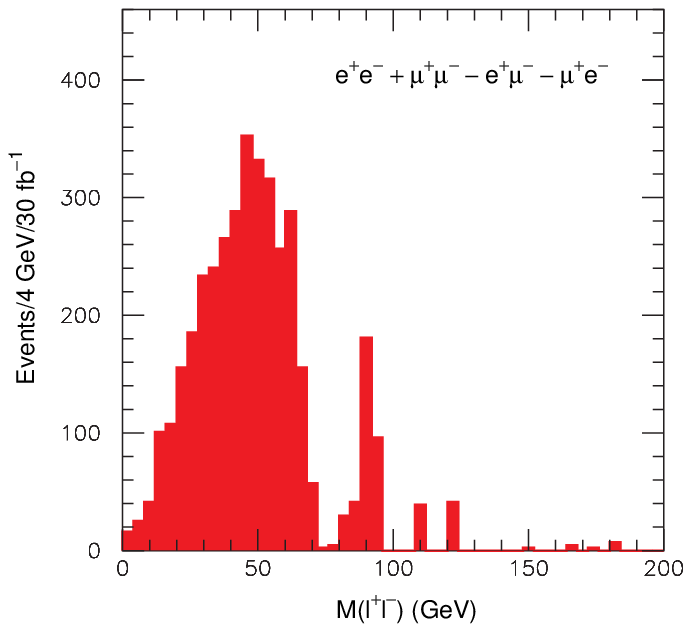}{p5_mllnew.epsi}
\caption{Examples of $M_{\ell\ell}$ for SUGRA points with $\tchi_2^0 \to \lsp
\ell^+\ell^-, \lsp Z$ (left) and for $\tchi_2^0 \to \tell_R^\pm
\ell^\mp$ (right).\protect\cite{TDR} \label{p4_mll}}
\end{figure}

\begin{figure}[t]
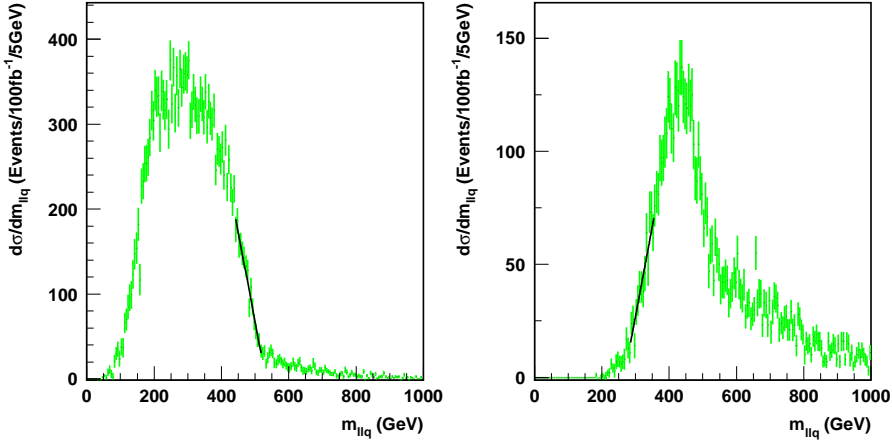

\dofig{\textwidth}{camb5sample.epsi}
\caption{Distributions of the smaller $M(\ell\ell j)$ (left) and the
larger $M(\ell\ell j)$ for $M_{\ell\ell} > M_{\ell\ell}^{\rm
max}/\sqrt{2}$ (right) mSUGRA ``Point 5.''\protect\cite{allanach}
\label{camb5new}}
\end{figure}

Long decay chains allow more endpoint measurements. The dominant source
of $\tchi_2^0$ at mSUGRA ``Point 5''\cite{TDR} and similar points is
$\tq_L \to \tchi_2^0 q \to \tell_R^\pm \ell^\mp q \to
\lsp\ell^+\ell^-q$. Assume the two hardest jets in the event are those
from the squarks and for each calculate $M(\ell\ell j)$, $M^{<}(\ell
j)$, and $M^{>}(\ell j)$.  Then the smaller of each of these should be
less than the endpoint $M_{\ell\ell q}$, $M_{\ell q}^{(>)}$, $M_{\ell
q}^{(<)}$ for squark decay, while the larger $M(\ell\ell j)$ should be
greater than the threshold $T_{\ell\ell q}$ requiring $M_{\ell\ell} >
M_{\ell\ell}^{\rm max}/\sqrt{2}$. These endpoints are smeared by jet
reconstruction, hadronic resolution, and mis-assignment of the jets that
come from squark decays.  Nevertheless, the distributions show clear
structure at about the right positions.

\begin{figure}[t]
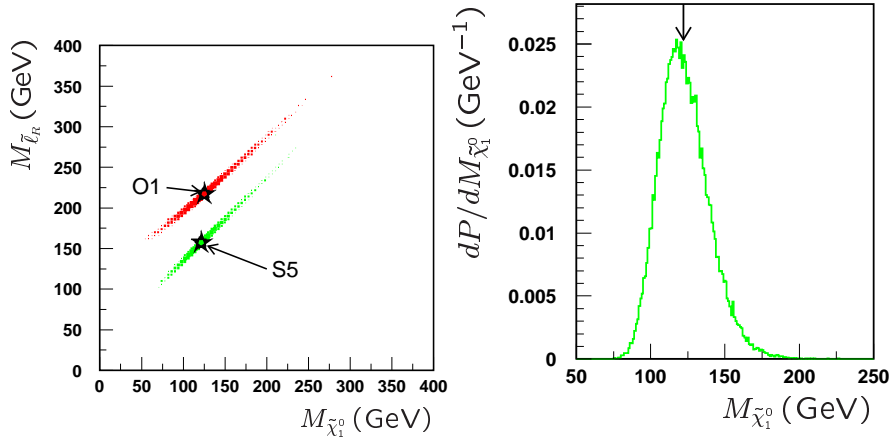

\dofigs{\dblfig}{camb15new.epsi}{camb11new.epsi}
\caption{left: Scatter plot of $M_{\tell_R}$ vs.\ $M_{\lsp}$ for two
models consistent with measurements described in the text. Right:
Projection of $M_{\lsp}$.\protect\cite{allanach} \label{camb15new}}
\end{figure}

After accumulating high statistics and careful study, it should be
possible to measure the endpoints to the expected hadronic scale
accuracy, $\sim 1\%$. The $\ell\ell q$ threshold is more sensitive to
hard gluon radiation, so it is assigned a larger error, $\sim 2\%$. Some
distributions of the resulting masses derived assuming these errors are
shown in Figure~\ref{camb15new} for two models, mSUGRA Point 5 (S5) and
an Optimized String Model (O1) with similar similar masses. Relations
among the masses are determined to $\sim 1\%$ and are clearly sufficient
to distinguish these models.  The LSP mass is determined to $\sim10\%$
by this analysis; since it is determined only by its effect on the
kinematics of the decay, the fractional error on $M_\lsp$ clearly
diverges as $M_\lsp^2 / M_\tq^2 \to 0$.

\section{$h \to b \bar b$ Signatures} 

If $\tchi_2^0 \to \lsp h$ is allowed, it may dominate over $\tchi_2^0
\to \lsp\ell\ell$. This signal can be reconstructed using two $b$ jets
measured in the calorimeter and tagged as $b$'s with the vertex
detector. A typical signal using the expected $b$-tagging efficiency and
light-quark rejection and the reach for such signals are shown in
Figure~\ref{p5_mbb}. Such a Higgs signal in SUSY events might well be
observed with less luminosity than $h \to \gamma\gamma$ or $h \to ZZ^*$
and so be the discovery channel for the light Higgs.

\begin{figure}[t]
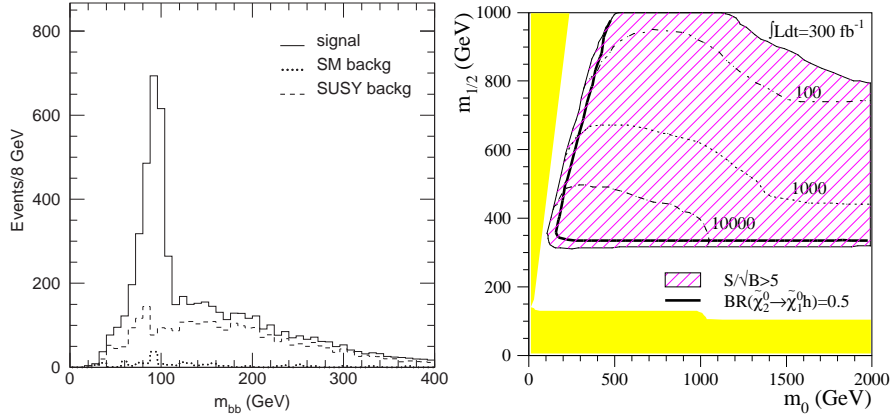

\dofigs{\dblfig}{p5_mbb.epsi}{tdr19-105.epsi}
\caption{Left: Typical signal for $h \to b \bar b$ in SUSY event sample
(``Point 5''). Right: Reach for this signal in mSUGRA.\protect\cite{TDR}
\label{p5_mbb}}
\end{figure}

If a signal for $h \to b \bar b$ is observed in SUSY events, the $h$ can
be combined with the two hardest jets in the event to measure the $\tq
\to \lsp h q$ endpoint in a way similar to the measurement of the
$\ell\ell q$ endpoint. The resulting distribution is shown in
Figure~\ref{p5_mbbj}; the endpoint is consistent with what is expected.
While the errors are worse than for the $\ell\ell q$ endpoint, the
measurement is still useful.

\begin{figure}[t]
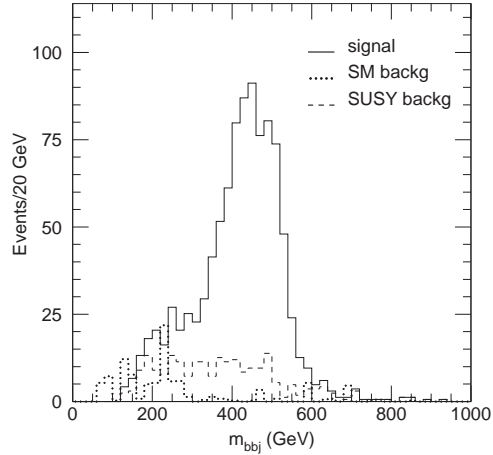

\dofig{\snglfig}{p5_mbbj.epsi}
\caption{$M_{hj}$ distribution for events with $h \to b\bar
b$.\protect\cite{TDR} \label{p5_mbbj}}
\end{figure}

\section{Heavy Gaugino Signatures}

In mSUGRA and other typical SUSY models, the light charginos and
neutralinos are mainly gaugino and so dominate the cascade decays, so
that
$$
B(\tq_L \to \tchi_2^0 q) \sim 1/3,\quad B(\tq_L \to \tchi_1^\pm q')
\sim 2/3, \quad B(\tq_R \to \tchi_1^0 q)\sim 1\,.
$$
But even in the simplest mSUGRA model, $\tchi_4^0$ and $\tchi_2^\pm$
have a significant admixture of gaugino and so contribute in light-quark
decays of squarks and gluinos. 

Four $\tchi_4^0/\tchi_2^\pm$ decay chains can give OS, SF dileptons:
$\tq_L \to \tchi^0_4 q \to \tell_{R}^{\pm} \ell^{\mp} q \to
\tilde\chi^0_2 \ell^+ \ell^- q$ [D1]; $\tq_L \tchi^0_4 q \to 
\tell_{L}^{\pm} \ell^{\mp} q \to \lsp \ell^+ \ell^- q$ [D2]; $\tq_L
\tchi^0_4 q \to \tell_{L}^{\pm} \ell^{\mp} q \to \tchi_2^0 \ell^+ \ell^-
q$ [D3]; and $\tq_L \to \tchi_2^\pm q' \to \tn_{\ell} \ell^{\pm} q'
\to \tchi^{\pm}_1 \ell^{\mp} q'$ [D4]. In principle these four decay
chains give four distinct $\ell^+\ell^-$ endpoints, but it seems
impossible to resolve these even with $100\,\fbi$ of integrated
luminosity. Nevertheless, there are $>10^3$ $\ell^+\ell^-$ events from
heavy gauginos over substantial range of mSUGRA parameters; $\tchi_4^0$
decays dominate for low $m_0$, while $\tchi_2^\pm$ dominates for the
region $m_0 \sim \mhalf$.

\begin{figure}[t]
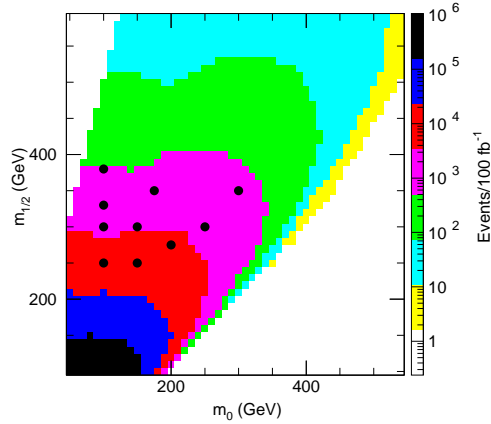

\dofig{\snglfig}{points.epsi}
\caption{Event rates for heavy gaugino decays in mSUGRA. The dots show
the 10 points studied in this analysis.\protect\cite{heavychi} 
\label{points}}
\end{figure}

\begin{figure}[t]
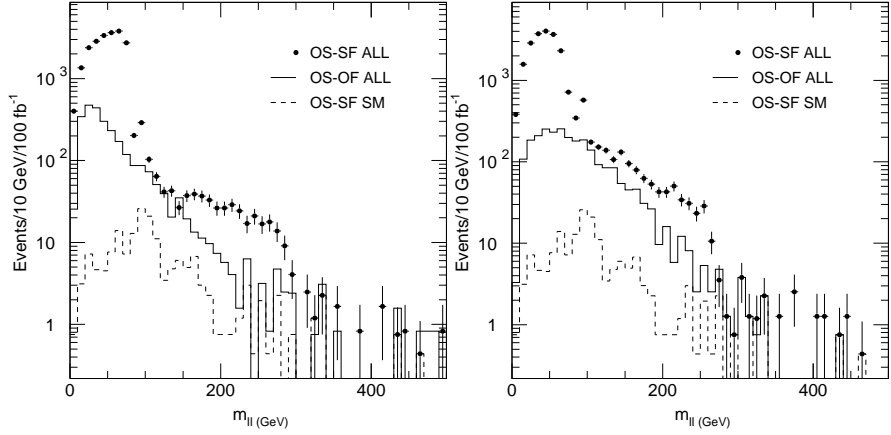

\dofigs{\dblfig}{sigloga.epsi}{sigloge.epsi}
\caption{$M_{\ell\ell}$ distributions for heavy chargino and neutralino
decays at $m_0,\mhalf = 100,250\,\GeV$ (left) and $150,250\,\GeV$
right.\protect\cite{heavychi} \label{sigloga}}
\end{figure}

\begin{figure}[t]
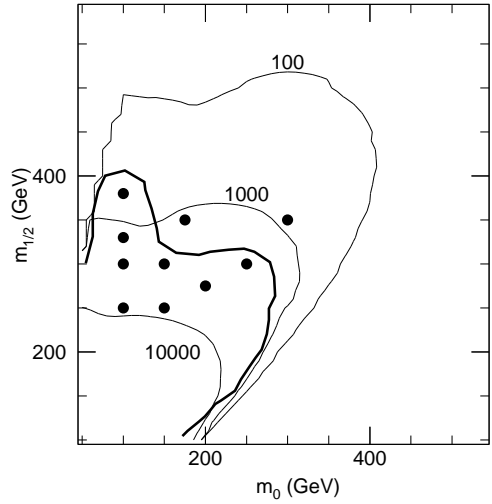

\dofig{\snglfig}{reach.epsi}
\caption{Reach in mSUGRA for heavy gaugino signals.\protect\cite{heavychi} 
\label{reach}}
\end{figure}

Event samples were generated and simulated for each of the ten points
indicated in Figure~\ref{points}. Events were required to have an
$\ell^+\ell^-$ dilepton pair, $M_{\ell\ell}>100\,\GeV$,
$\etmiss>100\,\GeV$, $\ge4$ jets, and $\Meff>600\,\GeV$. 
To suppress SM backgrounds a cut $M_{T2}>80\,\GeV$ was also made,
where\cite{t2}
$$
M_{T2}^2 \equiv \min_{\slashchar{p}_1+\slashchar{p}_2 = \slashchar{p}_T}
\left[ \max{ \left\{ 
m_T(p_{T,\ell_1},\slashchar{p}_1) , m_T(p_{T,\ell_2},\slashchar{p}_2)
\right\} } \right]
$$
is the minimum transverse mass obtained by partitioning the observed
$\etmiss$ between two massless particles. Note that $M_{T2} < M_W$ for
$t$ and $W$ backgrounds.

Results for $m_0,\mhalf = 100,250\,\GeV$ and $150,250\,\GeV$ are shown
in Figure~\ref{sigloga}. Evidently the signal is observable over the
SUSY and SM background in both cases. The estimated statistical error on
the endpoint is about $\pm4\,\GeV$ in both cases. The $5\sigma$ reach
for such signals in mSUGRA is indicated by the dark curve in
Figure~\ref{reach}.  Heavy gaugino signals are rather model dependent,
so the ability to study them is important for understanding the SUSY
model.

\section{Third-Generation Squark Signatures}

The properties of the third-generation squarks $\tilde b_{1,2}$ and
$\tilde t_{1,2}$ are important for understanding the SUSY model, but
their signatures are typically complex. The main production mechanism is
$\tg$ production and decay. Consider for mSUGRA with $m_0=100\,\GeV$,
$\mhalf=300\,\GeV$, $A_0=-300\,\GeV$, $\tan\beta=10$, $\sgn\mu=+$ the
processes
$$
\tg \to t \tilde{t}_1^* \to t \bar b \tchi_1^-, \qquad
\tg \to \bar b \tilde{t}_1 \to t \bar b \tchi_1^-
$$
Then the $M(t\bar b)$ endpoint can be used to measure a combination of
masses of the squark masses.

The analysis\cite{stop} requires as usual multiple hard jets and large
$\etmiss$ plus two jets tagged as $b$'s and two other jets $j$ not
tagged as $b$'s and consistent with $t \bar b \to jj b \bar b$. The
resulting $M_{t\bar b}$ distribution is still dominated by combinatorial
background. The next step is to select sidebands around $M_{jj}=M_W$,
rescale the jet momenta to $M_W$, and subtract to determine $t\bar b$
signal. The $M(t\bar b)$ mass distributions for one point before and
after subtraction are shown in Figure~\ref{hisano1}. The fitted endpoint
for this case is $443.2\pm7.4\,\GeV$ compared to expected $459\,\GeV$.
A similar agreement between reconstructed and expected endpoints was
found for all twelve points studied. Heavy squark signatures are clearly
difficult, but it appears possible to use a sideband analysis such as
this to study them with the ATLAS detector.

\begin{figure}[t]
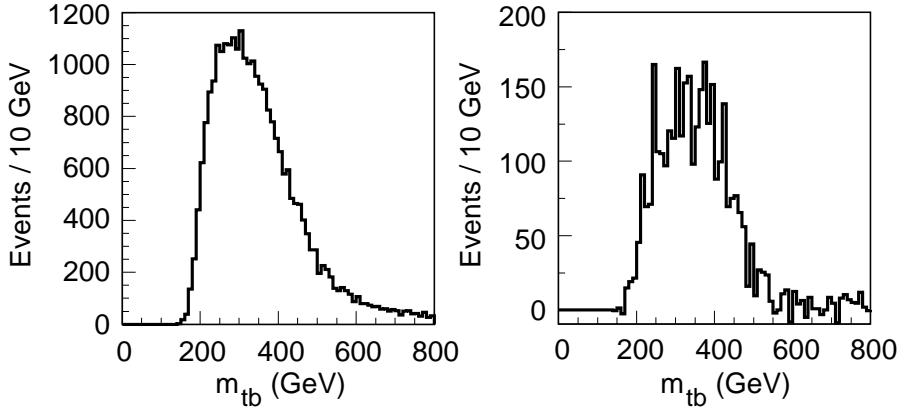

\dofig{\textwidth}{hisano1.epsi}
\caption{Reconstructed $M_{tb}$ distribution before sideband subtraction
(left) and after subtraction (right).\protect\cite{stop} \label{hisano1}}
\end{figure}

\section{$\tau$ Signatures} 

The mSUGRA model assumes $\tilde e$-$\tilde\mu$ universality, and this
is certainly suggested by the stringent limits on $\mu \to e\gamma$.
Even in the simplest mSUGRA model, however, the $\ttau$ behave
differently than $\tilde e$ and $\tilde\mu$ because of Yukawa
contributions to the RGE's, gaugino-Higgino mixing, and
$\ttau_L$-$\ttau_R$ mixing, which is $\propto m_\tau$. Hence $\tau$'s
provide unique information and might even be dominant in SUSY decays.

\begin{figure}[t]
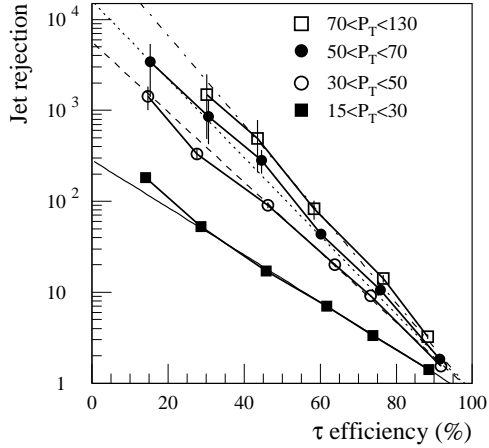

\dofig{\snglfig}{tdr9-31.epsi}
\caption{$\tau$ identification in ATLAS.\protect\cite{TDR} 
\label{tdr9-31}}
\end{figure}

The ATLAS (and CMS) vertex detectors cannot cleanly identify $\tau \to
\ell\nu\bar\nu$, so it is necessary to rely on hadronic $\tau$ decays.
The background for such decays is much larger than that for electrons
and muons. The $\tau$ efficiency vs.\ jet rejection shown in
Figure~\ref{tdr9-31} should be compared with the $>10^4$ rejection for
90\% efficiency expected for electrons and muons. Furthermore, all
$\tau$ decays contain missing neutrinos.  For $H,A \to \tau\tau$ one can
project $\etmiss$ on the measured $\tau$ directions to reconstruct the
$\tau\tau$ mass, but this is not possible for SUSY because of the
dominant $\etmiss$ from the $\lsp$'s.

\begin{figure}[t]
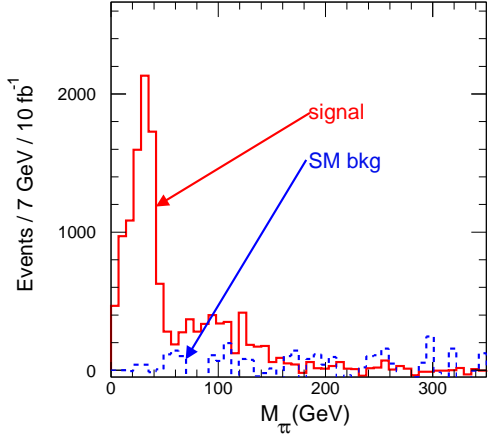

\dofig{\snglfig}{p6subtracted.epsi}
\caption{Reconstructed visible $\tau\tau$ mass in a mSUGRA event
sample.\protect\cite{TDR} \label{p6subtracted}}
\end{figure}

Decays into $\tau$'s are generally enhanced for $\tan\beta\gg1$. A
mSUGRA model with $m_0=\mhalf=200\,\GeV$, $A_0=0$, $\tan\beta=45$ gives
$\tchi_2^0 \to \ttau_1 \tau$ and $\tchi_1^\pm \to \ttau_1 \nu_\tau$ with
branching ratios close to unity. For events from this point, a simple
model for the detector response turns a sharp edge at $M_{\tau\tau} =
59.64\,\GeV$ into the distribution shown in Figure~\ref{p6subtracted}.
The visible momentum or mass depends both on the momentum and on the
polarization of the $\tau$. Measuring the $\tau$ polarization requires
separating different $\tau$ decay modes; the visible energy depends
strongly on polarization for $\tau \to \pi\nu$ but weakly for $\tau \to
a_1\nu$. Such a separation of decay appears to be possible, albeit
difficult: for example $\tau \to \pi\nu$ has a single track with $p = E$
and low electromagnetic energy. Recent work based on full GEANT
simulation has given an encouraging indication that the $\tau\tau$
endpoint can be inferred from the visible $\tau\tau$ mass.

\section{GMSB Signatures}

While the mSUGRA model remains after 20 years perhaps the most
attractive paradigm for SUSY breaking, it may not be correct. In the
GMSB model SUSY breaking is communicated via gauge interactions at a
scale much less than the Planck scale, so the gravitino $\tG$ is very
light. GMSB phenomenology\cite{gmsb} depends on the nature and lifetime
of NLSP ($\lsp$ or $\tell$) to decay into the $\tG$. In general the GMSB
model produces longer decay chains with more precisely measured decay
products,\cite{TDR} so reconstructing masses is considerably easier than
in mSUGRA.

\begin{figure}[t]
\dofigs{\dblfig}{tdr20-65.epsi}{tdr20-66.epsi}
\caption{Angular (left) and time (right) distributions for photons from
$\lsp \to \tG \gamma$ in ATLAS.\protect\cite{TDR} \label{tdr20-65}}
\end{figure}

\begin{figure}[t]
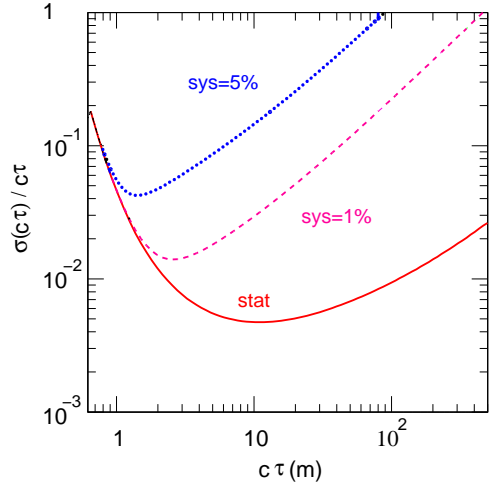

\dofig{\snglfig}{staulife.epsi}
\caption{Error on slepton lifetime for various assumptions on the
systematic error on the acceptance.\protect\cite{gpstau} \label{staulife}}
\end{figure}

The GMSB model can give a number of special signatures related to NLSP
decay. If the NLSP is a $\lsp$, its lifetime for $\lsp \to \tG \gamma$
can range from very short to very long. Short lifetimes can be detected
by using the Dalitz decays $\lsp \to \tG \ell^+ ell^-$ with branching
ratios of a few percent. Long lifetimes can be detected by looking for
(rare) non-pointing photons in SUSY events. The ATLAS electromagnetic
calorimeter has both good angular resolution in the polar angle, $\Delta
\theta \approx {60\,{\rm mr} \over \sqrt{E/1\,\GeV}}$, and good timing
resolution, $\Delta t \approx 100\,{\rm ps}$. Both can be used to detect
non-prompt photons from long-lived particles like $\lsp$ produced with
$\beta<1$. Such signals give a sensitivity up to $c\tau\sim100\,{\rm
km}$, much greater than what is expected in the GMSB model.

For other choices of the parameters the GMSB model might give long-lived
sleptons, which look like muons with $\beta<1$ in a detector. The ATLAS
muons chambers give a time-of-flight resolution in the $1\,{\rm ns}$
range over a distance of about $10\,{\rm m}$, making it possible to
reconstruct both the momentum and the mass of the slepton. The slepton
lifetime can be determined by comparing the rates for events with one
and two reconstructed sleptons as shown in Figure~\ref{staulife}. The
statistical error is small; the dominant systematic error is difficult
to estimate without real data. Another approach would be to look for
sleptons decaying into non-pointing tracks in the central detector. This
should be more sensitive for long lifetimes, but estimating the
sensitivity requires studying the pattern recognition for such
non-pointing tracks.

\section{Full Simulation of SUSY Events}

Most studies of SUSY signatures in ATLAS have been based on fast
simulation such as ATLFAST. While this should represent the ultimate
performance of the detector, it does not necessarily represent the
effort needed to achieve that performance. Therefore, a sample of 100k
SUSY events has recently been simulated with full GEANT for an mSUGRA
point with
$$
m_0=100\,\GeV,\ \mhalf=300\,\GeV,\ A_0=-300\,\GeV,\ \tan\beta=6,\
\sgn\mu=+
$$
The simulation of each event takes about $10^3\,{\rm s}$, compared with
about $1\,{\rm s}$ for event generation and fast simulation. Thus, such
a study represents a large effort.

Most of the effort so far has been devoted to debugging the
reconstruction software, so the results are not yet useful for assessing
the performance of the ATLAS detector for SUSY. However, a few physics
plots have been produced using cuts based on previous fast simulation
studies like those described above. As an example, Figure~\ref{lep_mll}
shows the $e^+e^- + \mu^+\mu^- - e^\pm\mu^\mp$ mass distribution from
full simulation of SUSY events after corrections for the average $e$ and
$\mu$ acceptance and for the $e$ energy scale. It is encouraging that
the distribution is quite similar to that obtained from fast simulation.

\begin{figure}[t]
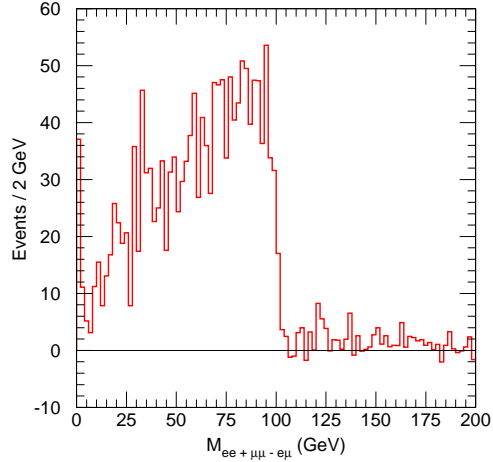

\dofig{\snglfig}{lep_mll.epsi}
\caption{$e^+e^- + \mu^+\mu^- - e^\pm\mu^\mp$ mass distribution from
full GEANT simulation after corrections for average acceptance and
electron energy scale.\protect\cite{athens} \label{lep_mll}}
\end{figure}

\section{Higgs Signatures}

SUSY requires Higgs bosons. In the Minimal Supersymmetric Standard Model
(MSSM) there are two Higgs doubles and hence after electroweak symmetry
breaking five Higgs bosons, $h$, $H$, $A$, and $H^\pm$. The light,
$CP$-even, $h$ satisfies $M_h<M_Z$ at tree level $M_h \simle 130\,\GeV$
after loop corrections. In many although not all SUSY models the $h$ is
very similar to a SM Higgs of the same mass.

The search for SM-like Higgs bosons has been a principle design goal of
both ATLAS and CMS, and a large amount of effort has been devoted to
studies of how to search for such particles. The global summary of these
studies is shown in Figure~\ref{higgs_01}: for each SM Higgs mass there
is at least one channel giving a significance of more than $5\sigma$
for an integrated luminosity of $100\,\fbi$, and the combined
significance of all channels is greater than about $10\sigma$.

\begin{figure}[t]
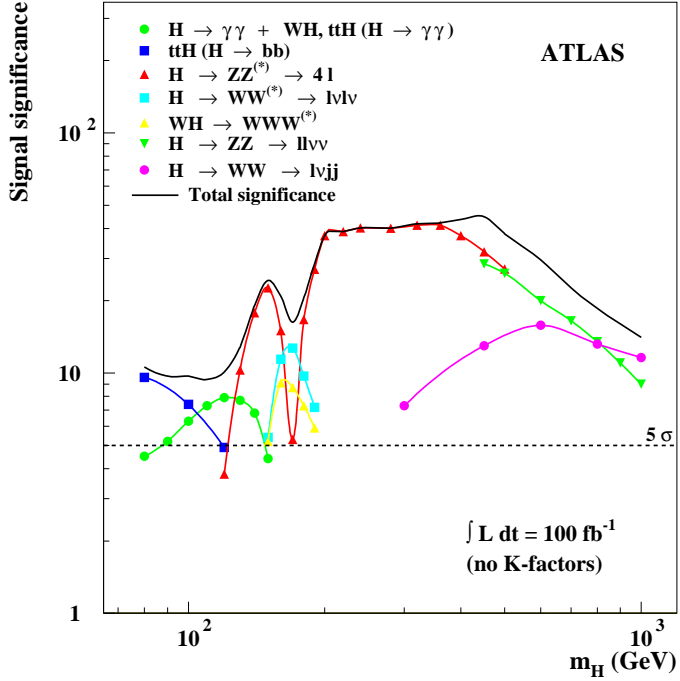

\dofig{3.5in}{higgs_01.epsi}
\caption{Significance of SM Higgs modes in ATLAS for
$100\,\fbi$.\protect\cite{TDR} \label{higgs_01}}
\end{figure}

Recently, more effort has been devoted to studies of how to measure the
properties of Higgs bosons once they are discovered. The key for doing
this is to observe the Higgs boson in more than one production and/or
decay channel.\cite{dieter} While $gg \to h$ is the dominant production
process at the LHC, $WW \to h$ is also significant and plays a crucial
role in the analysis. These events can be identified by requiring hard
forward jets resulting from the radiation of the $W$'s from incoming
quarks, $q \to W q'$, and no additional central jets. 

\begin{figure}[t]
\dofig{\snglfig}{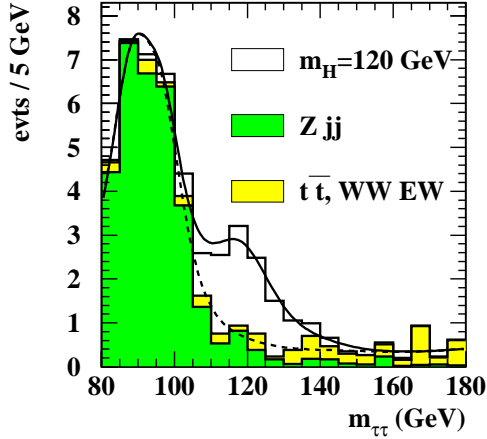}
\caption{Reconstructed $M_{\tau\tau}$ distribution from $WW \to h \to
\tau\tau$ after cuts.\protect\cite{jakobs} \label{m_tautau}}
\end{figure}

A typical result from such a study is shown in Figure~\ref{m_tautau}.
Events were selected using a combination of $e\mu$, $ee + \mu\mu$, and
$\ell h$ modes requiring a double forward jet tag and a central jet
veto; $M_{\tau\tau}$ was reconstructed by projecting $\etmiss$ on the
measured $\tau$ directions. The accepted cross section is about
$1.0\,\fb$ on a total SM background of $0.5\,\fb$. Thus this channel can
be used to measure the product of $\Gamma_{h,WW}$ and
$\Gamma_{h,\tau\tau}$. Combining a number of such measurements can give
a good determination of the properties of the $h$, although a linear
collider with sufficient energy and luminosity could do better.

\begin{figure}[t]
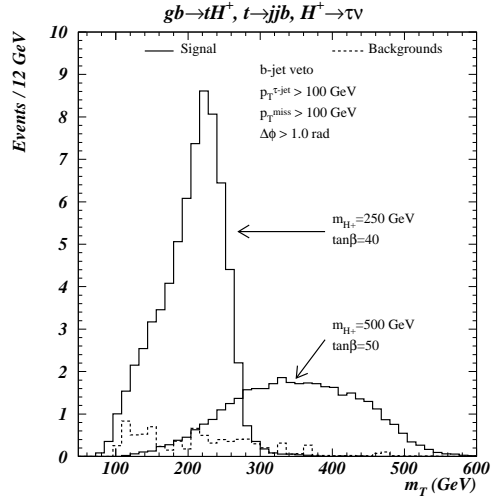

\dofig{\snglfig}{htnu5.epsi}
\caption{Transverse mass distribution for $gb \to tH^-$ with $t \to
q\bar q b$ and $H^- \to \tau^-\bar\nu$.\protect\cite{chiggs} 
\label{htnu5}}
\end{figure}

It may also be possible to reconstruct heavy Higgs bosons, especially
for $\tan\beta \gg 1$. An example of the reconstruction of $gb \to H^-
t$ with $H^- \to \tau^-\nu$ is shown in Figure~\ref{htnu5}. Since the
signature is a narrow hadronic $\tau$ plus reconstructed $t$, events
were selected requiring an identified hadronic $\tau$ jet plus two
non-$b$ jets and a $b$ jet consistent with $t$ kinematics. This analysis
relies on the fact that in $H^- \to \tau_R \nu \to \pi^- \nu\nu$ the
$\pi^-$ is hard and so is well separated from SM backgrounds.

\section{Outlook}

If SUSY exists at the TeV mass scale, ATLAS should find signatures for
it quite easily at the LHC. If $R$ parity is conserved, no mass peaks
for SUSY particles can be reconstructed, but several techniques have
been developed to measure combinations of SUSY masses using kinematic
distributions of observable decay products.

While the details of SUSY analyses at the LHC certainly depend on the
details of the SUSY model, it is possible to sketch a general outline of
how ATLAS could proceed first to search for and then to study SUSY with
$R$-parity conservation:
\begin{enumerate}

\item Search for an excess of multijet + $\etmiss$ events over the SM
expectation, and observe the $\Meff$ at which this emerges from the
background.

\item If such an excess is found, select a SUSY-dominated sample using
simple kinematic cuts.

\item Look in this sample for special features such as prompt $\gamma$'s
or long-lived $\tell$; either of these may occur in GMSB.

\item Look in the SUSY-dominated sample for $\ell^\pm$, $\ell^+\ell^-$,
$\ell^\pm\ell^\pm$, $b$ jets, hadronic $\tau$'s, etc.

\item Try simple endpoint-type analyses.

\end{enumerate}
Carrying out such an initial study seems quite feasible. Its results
would of course guide further more detailed and more model dependent
analyses.

\bigskip

I thank my many ATLAS collaborators who have contributed to the work
presented here. This work was supported in part by the United States
Department of Energy under Contract DE-AC02-98CH10886.

\end{document}